**Single Ion Implantation for Solid State Quantum Computer Development**

T. Schenkel[*], J. Meijer[1] and A. Persaud
E. O. Lawrence Berkeley National Laboratory, Berkeley, CA 94720

J. W. McDonald, J. P. Holder, and D. H. Schneider
Lawrence Livermore National Laboratory, Livermore, CA 94550

Several solid state quantum computer schemes are based on the manipulation of electron and nuclear spins of single donor atoms in a solid matrix. The fabrication of qubit arrays requires the placement of individual atoms with nanometer precision and high efficiency. In this article we describe first results from low dose, low energy implantations and our development of a low energy (<10 keV), single ion implantation scheme for $^{31}P^{q+}$ ions. When $^{31}P^{q+}$ ions impinge on a wafer surface, their potential energy (9.3 keV for $P^{15+}$) is released, and about 20 secondary electrons are emitted. The emission of multiple secondary electrons allows detection of each ion impact with 100% efficiency. The beam spot on target is controlled by beam focusing and collimation. Exactly one ion is implanted into a selected area avoiding a Poissonian distribution of implanted ions.

1. **Introduction**

Quantum computation holds the promise to revolutionize information technology [1-3]. Among the various fundamental implementation schemes of quantum computers, solid state approaches promise scalability to the large (>50) numbers of qubits required to outperform classical computers. In several widely discussed solid state quantum computer (SQC) proposals information is encoded in the spins of electrons and nuclei of phosphorous atoms in silicon or silicon-germanium hetero-structures [2, 3]. Phosphorous is the classical donor element in the Si IC industry. In a SQC, the excess electron is not ionized and dopant atoms are spaced with distances of 20 nm (silicon) to over 100 nm (silicon germanium hetero-structures) so that their wave functions can overlap. Single and two qubit operations are performed with gate electrodes and using single electron transistors for readout through spin conditional charge measurements. Scanning tunneling microscopy [4-6] and single ion implantation [6-8] are two basic approaches for the fabrication of $^{31}P$ qubit arrays. In this article, we describe our program in the development of single ion implantation technology for solid state quantum computer fabrication. In the following, we first discuss the basic problem of ion implantation in respect to the stringent requirements on implantation profiles and process integration posed by SQC designs. We then describe in some detail the single ion implantation system that is currently being built at LBNL.

2. **Low dose, low energy implantation of $^{31}P$ ions into silicon**

In the SQC scheme proposed by Kane, $^{31}P$ atoms are spaced with a period of about 20 nm and at a depth of about 10 nm. The equivalent ion dose is 2.5E11 cm$^{-2}$. The kinetic energy of ions sets the mean penetration depth in a given matrix. Ions can be implanted through a dielectric barrier layer into the required depth of 10 nm, or more shallow implants with lower impact energies can be followed by thin film deposition steps [3]. The kinetic energy of dopant ions will be below 10 keV, a physical lower limit is set by the need to incorporate the dopant a few monolayers deep in the matrix, while a technological limit on the implant energy results from focusing properties in the implanter and the single ion detection scheme used. Ion implantation for SQC fabrication is in the low dose and low energy regime.

Ion implantation is not a gentle process. Ions impinge on silicon wafers with kinetic energies sufficiently large to displace silicon atoms from their lattice positions, forming vacancies and possibly extended defects. Impinging ions deposit their kinetic energy in statistical energy loss processes resulting in range profiles with lateral and longitudinal distributions. When dopant atoms come to rest, they mostly do not occupy lattice positions and are not electrically active. Electrical activation of dopants and repair of damage to the matrix crystal are achieved through annealing. During annealing, dopant atoms inevitably diffuse. In this low dose regime this is due to intrinsic transient enhanced diffusion (TED) mechanisms [9]. Diffusion jeopardizes a carefully defined array structure. Annealing parameters, i. e., temperatures in the range of 600 to 1100 C° and annealing durations (in the range of a few seconds to minutes) required for full (>95%) electrical activation and matrix repair depend on the ion species, dose and kinetic energy. For $^{31}$P on Si at 1 to 10 keV and doses below 1E12 cm$^{-2}$, the literature is sparse, or non existent.

---

[*] e-mail: T_Schenkel@LBL.gov

We are currently developing an SQC processing protocol including annealing conditions for low dose, low energy P implants in Si with the goal to optimize electrical activation and damage repair while keeping diffusion at a minimum. In this study we use p-type silicon (100) test wafers with a resistivity of ~80 Ohm cm. Wafers are implanted with $^{31}$P$^{1+}$ ions at an incident angle of 7° off normal. Off axis implantation reduces channeling. The native oxide of ~1.5 nm thickness was not stripped prior to implantation and functioned as a thin screen oxide. Dopant ions were placed randomly into the silicon wafers. The implantation dose was set to 2.5E11 cm$^{-2}$. The depth distribution of dopant atoms is measured with Secondary Ion Mass Spectrometry (SIMS). We used a magnetic sector instrument (CAMECA 4f) to ensure separation of secondary ion signals from $^{31}$P$^{+}$ and $^{30}$SiH$^{+}$. Probe beams were 3 keV O$_2^+$ with an incident angle of 52° with oxygen leak. Quadruple SIMS might offer lower primary ion beam energies and thus slightly better depth resolution, but the mass resolving power is insufficient to exclude interference from silicon hydride. Depth profiles for implants with 2.5 keV and 5 keV energies are shown in Figure 1 together with data from a blank control sample.

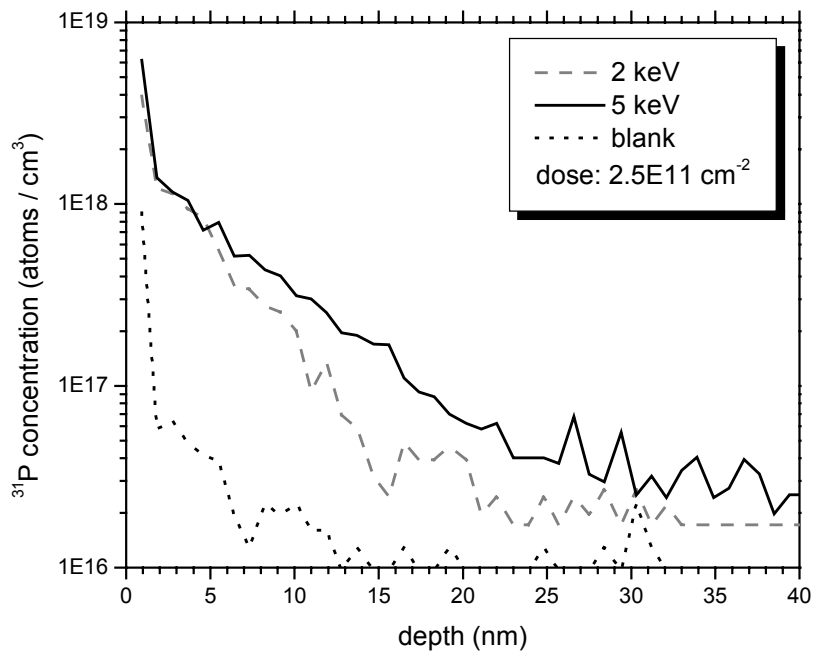

Figure 1: Magnetic sector SIMS depth profiles of low dose, low energy $^{31}$P implants in silicon.

The depth profiles were take from as implanted wafers, prior to oxide stripping and annealing. The high concentration in the oxide stems in part from ambient contamination. Repeated measurements of blank samples showed a surface near phosphorous concentration of about 8E17 atoms/cm$^3$. Also, the data are not corrected for the transient in secondary ion formation probability at the beginning of the dynamic SIMS analysis, when sample and beam are not yet equilibrated and the transient due to the crossing of the silicon dioxide-silicon interface. The two impact energies result in very similar dopant profiles, with peak concentrations of 1.4E18 atoms/cm$^3$ in a depth of 2 nm, just below the native oxide. The 5 keV implant shows a more pronounced tail below a depth of 5 nm. This tail probably contains a channeling contribution [10]. It is apparent that a significant fraction (up to 50%) of dopant atoms will be lost when implanting through even a few nm thin layer of resist or a dielectric barrier. The fraction lost in a such a layer in the implantation process can be reduced by increasing the implantation energy, but this also increases the lateral and longitudinal straggling of dopant profiles.

Phosphorous diffuses with silicon interstitials and the effect of diffusion retardation due to injection of vacancies from a thin silicon nitride barrier layer might reduce or limit intrinsic TED during annealing. On the contrary, the presence of an $SiO_2$/Si interface during annealing leads both to enhanced loss of electrically active dopants and enhanced diffusion [9]. Recently, the effect of dopant spacing on interdopant coupling through the exchange interaction in an SQC configuration was calculated by Koiller et al [11]. Clearly, the lateral and longitudinal width of dopant profiles will lead to a fraction of inactive qubit sites. Straggling can be minimized by minimizing the implantation energy. Dopant loss in a dielectric barrier layer can be excluded by implantation into clean silicon. Both of these measures have other adverse consequences. Very low energy implantation into clean silicon (or $Si_xGe_y$) with alignment of single ion implants to sacrificial marker structures, followed by annealing and deposition of a barrier layer [3] resembles a hybrid between the STM based bottom up approach and the top down approach of ion implantation through a barrier layer into the matrix crystal [6]. Clearly, qubit yield management is crucial in the development of workable device processing schemes.

## 3. Single ion implantation with highly charged $^{31}$P ions

The formation of qubit arrays of individual dopant atoms has to solve two challenges. First, ions have to be implanted into defined areas and with spacings of 10 to 100 nm. Second, only one ion should be implanted into each area. Our approach uses a focused and collimated beam of highly charged $^{31}P^{q+}$ (q=10 to 15+) ions. Highly charged ions are extracted from an Electron Beam Ion Trap (EBIT) [12]. We routinely extract $P^{q+}$ beams with beam currents of about 1E7 $^{31}P^{12+}$/s and have recently transported a few thousand $^{31}P^{12+}$/s through a 5 μm aperture [13].

When slow, highly charged $^{31}P^{15+}$ ions impinge on solid surfaces, their potential energy is released and about 20 secondary electrons are emitted. The emission of secondary electrons is largely independent of the impact energy in the range from 1 keV to 100 keV [14]. The high secondary electron yields enable the detection of each ion impact with 100% efficiency. Detection of each ion then allows implantation of exactly one ion into a selected area, and a Poissonian implant distribution can be avoided. This single ion implantation scheme is independent of the target material, and independent of the ion impact energy.

The deposition of potential energy also leads to very efficient resist development by individual ion impacts. This was demonstrated in atomic force microscope studies of defects from single ion impacts with $Xe^{41+}$ ions and self assembled alkyl monolayers on Si (111) [8].

In Figure 2, we show a schematic of the single ion implantation setup that is currently constructed at LBNL. Beams of highly charged $^{31}P^{q+}$ ions are analyzed in a Wien filter and ions of a selected charge state enter the acceleration – deceleration lens. The beam is focused electrostatically and then collimated in an aperture with a diameter of 10 to 50 nm. This last aperture defines the spot size on target. Apertures are currently developed based on chemically assisted focused ion beam milling of 200 nm thick low stress silicon nitride membranes with a 30 keV $Ga^+$ beam (Figure 3). The single ion implantation system will allow *in situ* imaging of sample structures by SEM and alignment of implant positions to pre-defined structures.

**Ion optical calculations**

The requirements for an ion optical system that can achieve beam spots below 100 nm at energies below 10 keV are very high. Especially because the brightness of the EBIT is much smaller than that of liquid metal ion sources [15]. The energy spread has been found to be in the range of ~10 eV. In the following, we use parameters described by Marrs et al. [15]. Here, the ion current is about I=$10^6$ ions/s for a beam spot with r = 0.3-0.4 mm at a divergence of r´=3 mrad and an energy of 17 keV per ion charge. Follow the Liouville theorem

$$I = \beta \cdot (r \cdot r')^2 \cdot E \qquad (1)$$

with a constant β results in the requirement that the divergence is 469 mrad for E=100 eV per ion charge and a 50 nm beam spot at a current of 1 ion/s. This leads to an expected focus depth of such a system of only 50 nm. The additional high chromatic aberrations and the expected parasitic distortion like vibrations or the influence of residual alternate magnetic fields makes a technical solution using conventional focussing systems very difficult.

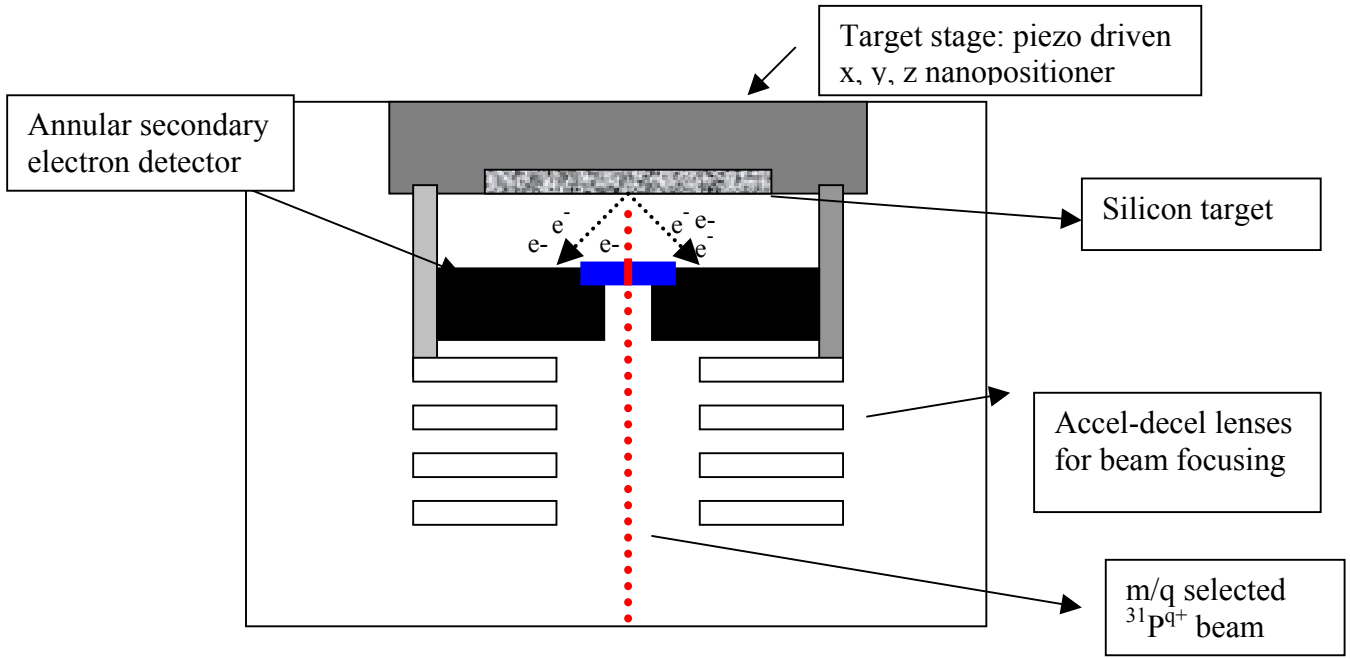

Figure 2.: Schematic of the SII system.

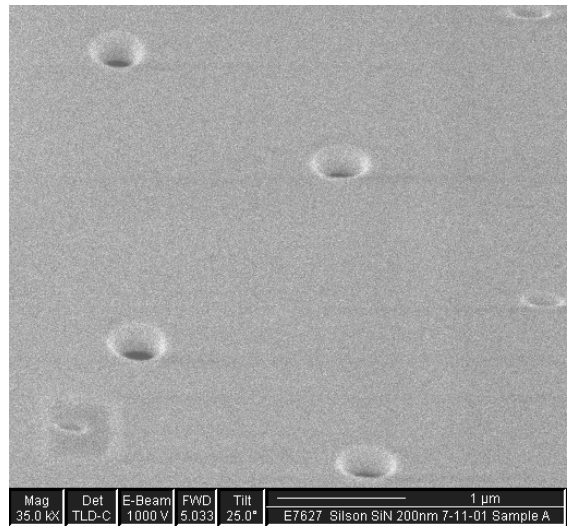

Figure 3: Example of test apertures from FIB drilling in thin SiN membranes.

One possible ion optical design solution of this problem is the illumination of a mask near the surface of the target or the use of micro lenses. Micro lenses are attractive but have the problem of limited incoming divergence and the resolution requirements make lens fabrication a complicated micro mechanical task.

Beam collimation is usually not reasonable due to significant slit scattering of high energy ions. For low energy ions, slit scattering is negligible. If the mask or aperture is position near the surface the acceptance divergence angle could be very high. Now the Liouville theorem shows us that in principle these solution should work, but the ion optical problem still exist and the question is how to build a system that allows to focus the beam to a smallest spot size at the lowest possible energies. A desired spot size of 50 nm at a current of 1 ion/s leads to a required beam spot of 10 µm with a current of more than $4 \cdot 10^4$ ions/s at the aperture or stencil mask. We performed ray tracing calculations using a self written code called SCP [16]. The program allows a three dimensional calculation of the potentials using a finite difference method (FDM). In order to reduce the calculation time FDM is combined with a multigrid and over-relaxation method. The subsequent ray tracing technique is used for an calculation of the smallest beam waist and the influence of different parasitic aberrations to the beam spot, like the earth magnetic field. Through variation of beam parameters the program allows a fast optimisation of the beam spot size. Additionally, we compared results with simulations from the widely used SIMION code.

The strategy to find an optimised solution was first to accelerate the beam so as to reduce the divergence and the beam spot size while the ions travel though the focussing lens system and then stop the beam as quickly as possible to reduce the time where aberrations (especially the chromatic aberration) affect the beam spot size. Additionally, the system should allow easy tuning of only one element to find the focal spot. A schematic of the lens layout and the potential distribution is shown in fig. 3.

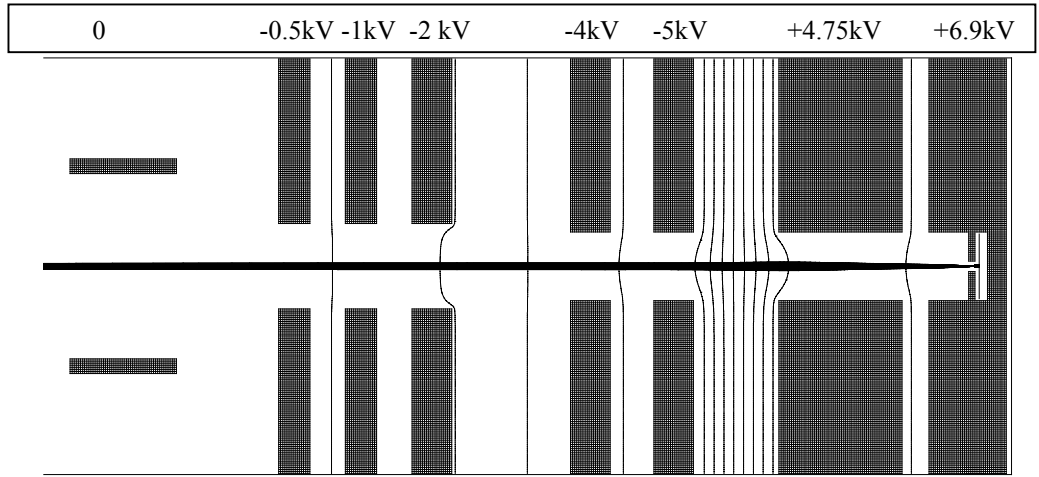

Figure 3: Schematic of lens geometry with potential distributions and beam trajectories from ray trace calculations with SIMION.

The asymmetric accelerator/decelerator lens system is optimised to decelerate the ions from 7 keV × q to 0.1 keV × q. The demagnification of the system is 15.2 and the magnification of the divergence is found to be 108, enlarged by $\sqrt{E}$ due to the deceleration. A chromatic aberration of this system is found to be $\Delta x = C_c \cdot r' \cdot \frac{\Delta E}{E}$ with $C_c$= 0.1 [m/rad] and a spherical aberration coefficient given by

$\Delta x = C_s \cdot (r')^3$ is found to be $C_s$=0.008 [m/rad$^3$]. We find a beam current of several ions/s at a beam spot size of 50 nm for final beam energies between 0.15 and 1 keV×q.

The simulations described here use as the take off point ion source and beam parameters from Marrs et al. [15]. Since then the ion source performance has been optimised for $^{31}$P$^{q+}$ production. We routinely extract ion beams with intensities at the 1E6 ions/s level, improvement to 1E7 ions/s is anticipated. Optimisation of ion source parameters has the potential to reduce the energy spread and increase the brightness at a given current level by an order of magnitude [17].

### 4. Conclusions

We have presented a basic discussion of the ion implantation problem associated with the formation of qubit arrays for solid state quantum computer schemes. Qubit yield limiting factors (such as dopant loss in a barrier layer) have to be quantified in the definition of workable device processing flows. The single ion implantation scheme with $^{31}$P$^{q+}$ ions is described and ion optical calculations are presented that outline the basic principles of this approach. While significant challenges remain, we conclude that single ion implantation is a key enabling technology for the realization of solid state quantum computers.

[1] permanent address: Department of Physics, Ruhr University Bochum, Germany

**Acknowledgments**

We thank Jeff Kingsley and Ming Hong Yang from Charles Evans and Associates, Sunnyvale, CA, for the FIB and SIMS work, respectively.

This work was supported by the U. S. Department of Energy under Contract No. DE-AC03-76SF00098 and by the National Security Agency (NSA) and Advanced Research and Development Activity (ARDA) under Army Research Office (ARO) contract number MOD707501. Work at Lawrence Livermore National Laboratory was performed under the auspices of the U. S. Department of Energy under contract No. W-7405-ENG-48.